\documentclass[epj-spec]{svjour}

\usepackage{graphicx,color,amssymb}

\newcommand{\lbco}     {${\rm La}_{1.875} {\rm Ba}_{0.125} {\rm Cu O_4}$}
\newcommand{\slrr}      {$T_1^{-1}$}

\newcommand{\lesxco}    {${\rm La}_{1.8-x} {\rm Eu}_{0.2} {\rm Sr}_{x} {\rm Cu O_4}$}

\newcommand{\lsxco}     {${\rm La}_{2-x} {\rm Sr}_{x} {\rm Cu O_4}$}
\newcommand{\lbcox}     {${\rm La}_{2-x} {\rm Ba}_{x} {\rm Cu O_4}$}

\newcommand{\la}        {$^{139}$La}
\newcommand{\cu}        {$^{63,65}$Cu}
\newcommand{\latone}      {$^{139}T_1^{-1}$}

\newcommand{\otone}      {$^{17}T_1^{-1}$}

\newcommand{\oxy}       {$^{17}$O}

\usepackage{color}
\definecolor{orange}{rgb}{0.7,0.6,0.1}
\definecolor{turkis}{rgb}{0,0.7,0.3}

\begin{document}

\title{Charge order and low frequency spin dynamics in lanthanum cuprates revealed by Nuclear Magnetic Resonance}

\author{H.-J. Grafe\inst{1} \and N. J. Curro\inst{2} \and B. L. Young\inst{3} \and A. Vyalikh\inst{1}\thanks{Present address: Leibniz-Institute of Polymer Research Dresden, Hohe Str. 6, 01069 Dresden, Germany} \and J. Vavilova\inst{1,5} \and G. D. Gu\inst{4} \and M. H\"ucker\inst{4} \and B. B\"uchner\inst{1}}

\institute{IFW Dresden, Institute for Solid State Research, P.O. Box 270116, D-01171 Dresden, Germany \and Department of Physics, University of California, Davis, Ca 95616, USA \and Department of Electrophysics, National Chiao Tung University, Hsinchu 300, Taiwan \and Condensed Matter Physics \& Materials Science Department, Brookhaven National Laboratory, Upton, NY 11973-5000, USA \and Kazan Zavoiskiy Physical-Technical Institute, Kazan, Russia}

\date{Received: date / Revised version: date}

\abstract{We report detailed \oxy , \la , and \cu\ Nuclear Magnetic Resonance (NMR) and Nuclear Quadrupole Resonance (NQR) measurements in a stripe ordered \lbco\ single crystal and in oriented powder samples of \lesxco . We observe a partial wipeout of the \oxy\ NMR intensity and a simultaneous drop of the \oxy\ electric field gradient (EFG) at low temperatures where the spin stripe order sets in. In contrast, the \cu\ intensity is completely wiped out at the same temperature. The drop of the \oxy\ quadrupole frequency is compatible with a charge stripe order. The \oxy\ spin lattice relaxation rate shows a peak similar to that of the \la , which is of magnetic origin. This peak is doping dependent and is maximal at $x$ $\approx$ 1/8.}

\titlerunning{Charge order and low frequency spin dynamics in lanthanum cuprates revealed by NMR}
\maketitle

\section{Introduction}
\label{intro}
The high temperature superconducting cuprates (HTSC) are well known for their unusual magnetic and electronic normal state properties. Many of these properties are believed to play an important role in the mechanism of high temperature superconductivity. The intrinsic inhomogeneous distribution of charges and spins in the CuO$_2$ planes of the HTSC's is a particular phenomenon that has attracted widespread attention. The discovery of the anomalous suppression of superconductivity in \lbcox\ with $x$ = 1/8, and in rare-earth co-doped ${\rm La}_{2-x-y} {\rm RE}_{y} {\rm Sr}_{x} {\rm Cu O_4}$ suggests that the inhomogeneity is not random in these materials, but that the charges collect into one-dimensional stripes that are separated by hole poor, antiferromagnetically (AF) ordered regions. Evidence for such a charge order has been found by diffraction techniques such as Neutron scattering \cite{tranquada1,tranquada2,fujita} and x-ray diffraction \cite{abbamonte,fink,kim,kim2,zimmermann,hucker}. A recent combined Angle Resolved Photo Emission Spectroscopy (ARPES) and Scanning Tunneling Microscopy (STM) study is consistent with these results \cite{valla}. More  indirect methods such as optical conductivity and transport and magnetization properties also show anomalous effects at the onset temperature of charge order \cite{homes,li,tranquada3,hucker2}. 

Nuclear Magnetic Resonance is a local probe and can therefore not directly distinguish a stripe order from a random or other spin-charge orders. But there are many NMR studies that provide (i) evidence for inhomogeneities in the CuO$_2$ planes \cite{singer,grafe,haase2,singer2}, and (ii) evidence for slowly fluctuating electronic spins that mark the onset of the spin stripe order \cite{hunt,nick,julien,teitelbaum,imai,barbara2,grafe}, for a review see Ref. \cite{imai} and \cite{imai2}. In this article we report \oxy\ and \la\ NMR, and \cu\ NQR results on stripe ordered \lbco\ (LBCO), and compare these results with measurements on \lesxco\ (LESCO). Our primary result is the discovery that the planar oxygen EFG, $\nu_c$, becomes temperature dependent at low temperatures where the stripe order as determined by other methods sets in. At the same temperature, the intensity of the planar oxygen signal is partially wiped out, and the intensity of the Cu NQR signal is completely lost. These effects are possibly due to different couplings of the Cu and O nuclei to the electronic spin system, and the results presented here directly tie in with the results on \lesxco\ presented in a previous article \cite{grafe}. Furthermore, we show here that we can distinguish the apical and the planar oxygen spectra at higher temperatures. We have measured the spectra with different repetition times and different orientations of the external magnetic field, and find that at low temperatures a signal with a reduced EFG remains. Due to intensity arguments and the fact that the quadrupole frequency of the apical oxygen is higher than that of the planar oxygen at high temperatures, we relate the decrease of the quadrupole frequency at low temperatures to the planar oxygen spectra. In addition, we have measured the spin lattice relaxation rate, \slrr\ , of the oxygen and the lanthanum. At low temperatures, \slrr\ of both \la\ and \oxy\ shows a peak that is due to the slowing down of electronic spin fluctuations. The peak occurs at the same temperatures for lanthanum and oxygen, confirming the magnetic origin of the peak, and shows a pronounced doping dependence: it is maximal and occurs at higher temperatures for $x$ $\approx$ 1/8. For higher dopings, the amplitude of the peak is lower, and it occurs at lower temperatures. For lower dopings, the peak is of the same height as for $x$ $\approx$ 1/8, but occurs at lower temperatures.

\section{Experiment}

\begin{figure}
\begin{center}
 \includegraphics[width=90mm,clip]{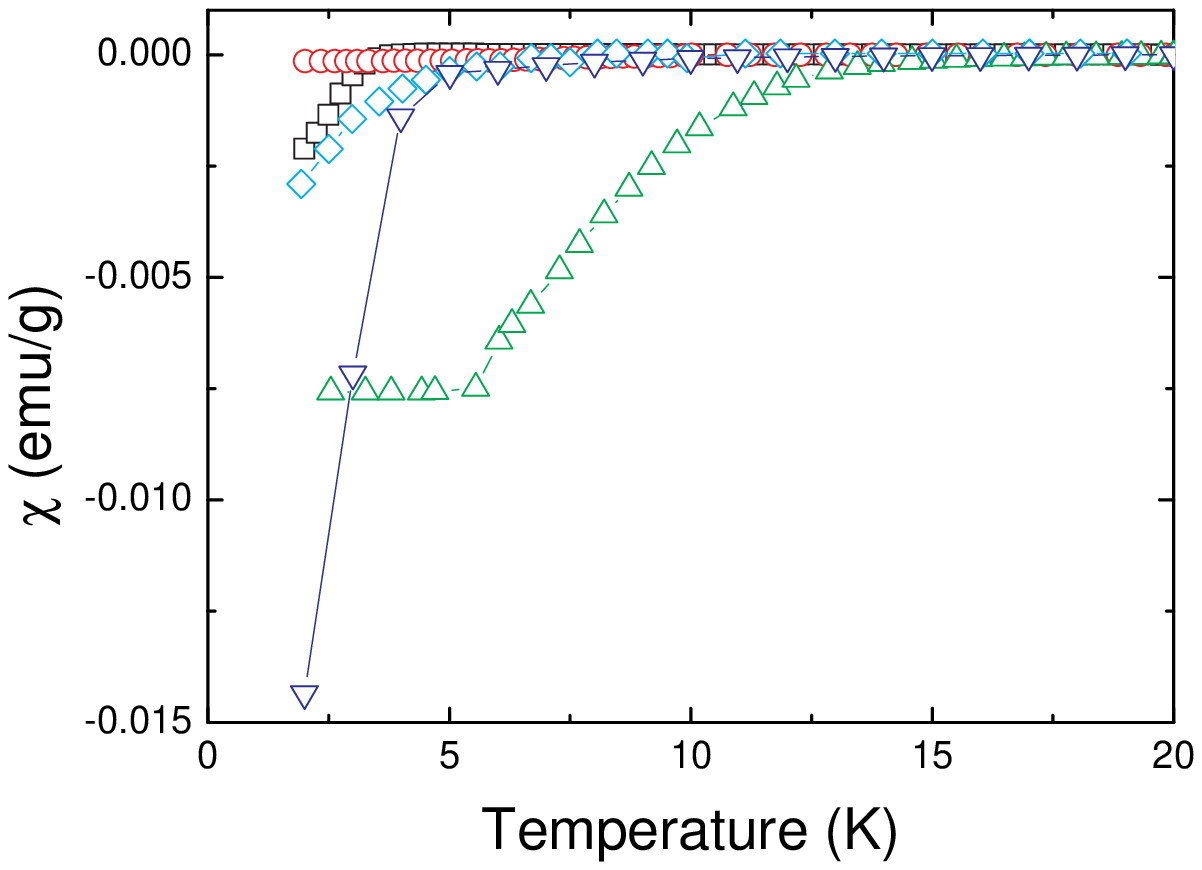}
 \caption{(Color online) Zero-field-cooled (ZFC) susceptibility of \lbco\ single crystal measured at 20 G and \textbf{H}$||c$. ($\square$) measured as is after preparation, ($\textcolor{red}{\circ}$) after heating in $^{16}$O$_2$ atmosphere at 850 $^{\circ}$C for 96 h, ($\textcolor{green}{\triangle}$) after heating in $^{17}$O$_2$ atmosphere at 850 $^{\circ}$C for 10 days, ($\textcolor{blue}{\triangledown}$) after second and third heating in $^{17}$O$_2$ atmosphere at 400 $^{\circ}$C for 48 h and 96 h, respectively. ($\textcolor{cyan}{\diamond}$) reproduced data from Fujita \textit{et al.} \cite{fujita} measured at 10 G.}
 \label{susce}
\end{center}
\end{figure}

We have measured a single crystal of LBCO with a Ba content of $x$ = 0.125. The sample preparation is described elsewhere \cite{fujita}. For the \oxy\ NMR measurements the crystal was enriched by annealing in $^{17}$O$_2$ gas at different temperatures. The first enrichment at 850 $^{\circ}$C changed the oxygen content of the crystal so that it became superconducting at low temperatures. The superconductivity was observed in a SQUID
magnetometer by measuring the susceptibility in an external field of 20 G. The temperature dependence of the susceptibility is shown in Fig. \ref{susce}.
The same crystal also exhibited no temperature dependence of the electric field gradient (EFG), as measured by \oxy\ NMR, suggesting that the charge deficiency introduced by the oxygen defects created at 850 $^{\circ}$C not only gave rise to the superconductivity, but destroyed the static inhomogeneous charge ordering. The crystal was then annealed again at a temperature of 400 $^{\circ}$C in \oxy$_2$ atmosphere. Eventually, only traces of superconductivity could be observed at temperatures below 4 K. As we see below this sample showed characteristic effects of the
stripe order in the NMR measurements.

Zero-field Cu NQR was measured in LBCO and LESCO, $x$ = 0.13, to check the Cu wipeout. The Cu NQR frequency spectra were obtained by integration of Cu nuclear spin echoes with changing the frequency.

The \oxy\ NMR spectra were taken at a fixed frequency of $f$ = 53.124 MHz by sweeping the magnetic field. We used a standard Hahn-Echo sequence, and integrated the echo at each field step. A tuned NMR/NQR circuit with an almost temperature independent low quality factor $Q$ similar to that in Ref. \cite{gvmw} was used to minimize the influence of $Q$ on the spectral intensity $^{\rm O}I_{\rm NMR}$. The measurements on the \lesxco\ were done
on the same aligned polycrystals that were used in our previous study \cite{grafe}, with Sr contents $x$ = 0.08, 0.105, 0.13, 0.17, and 0.2. The resonance field of each transition at fixed frequency $f$ is given by $\gamma H_n = (f-n\cdot\nu_c)/(1+K_c)$ , where $n$ = -2, -1, 0, 1, or 2, and $K_c$ is the Knight shift. The spectra were fit to Lorentzian distributions centered at the $H_n$ with widths $\sigma =  \sqrt{\sigma_m^2+(n\cdot\sigma_q)^2}$, where $\sigma_m$ is the magnetic linewidth and $\sigma_Q$ the quadrupolar linewidth caused by the distribution of $\nu_c$. The spin lattice relaxation rate, $T_1^{-1}$, of the \oxy\ and the \la\ was measured by inversion recovery, where an additional $\pi$
pulse is applied at a delay time $t$ before the usual Hahn-Echo sequence. We integrated the echo for different delay times $t$ and fit the data to different expressions as described below.

\oxy\ NMR is ideal for probing the stripe order in the cuprates, because the oxygen nucleus has a spin of $I$ = 5/2, and a quadrupole moment of -2.558 barn. The nuclear spin interacts with the electronic spins and orbital moments of the Cu and O orbitals, whereas the quadrupole moment interacts with the EFG of the crystal. The quadrupolar interaction acts to split the 5/2 multiplet into five different lines separated by $\nu_c$, where $\nu_c =3eQV_{cc}/2I(2I-1)h$ is a measure of the EFG at the O site, $V_{cc}=\partial^2V/\partial x_c^2$ is the $cc$ component of the EFG tensor, $e$ is the electron charge and $h$ is Planck's constant. The EFG has two contributions: a lattice component, and an on-site component arising from partially unoccupied orbitals. For example, a hole in the planar oxygen $p$ orbital ($2p^5$ configuration) gives a substantial contribution to the EFG, whereas an isotropic distribution of charge ($2p^6$ configuration) produces no EFG at the oxygen nucleus. \cite{takigawa,haase} The EFG of the lattice comes from distant ions and their deforming effect on closed shells. This lattice contribution does not depend on doping. Therefore, it has been found that $\nu_c$, is directly proportional to the hole doping, $n_h$, in \lsxco.\cite{haase} We find a similar trend in \lesxco\ at high temperatures\cite{grafe}, where additional Eu doping does not change the slope of $\nu_c$ versus $x$, and we infer that a similar relationship exists for LBCO as well.

\section{Results and Discussion}

\subsection{\oxy\ NMR spectra of \lbco\ and EFG }

\begin{figure}
\begin{center}
 \includegraphics[width=70mm,clip]{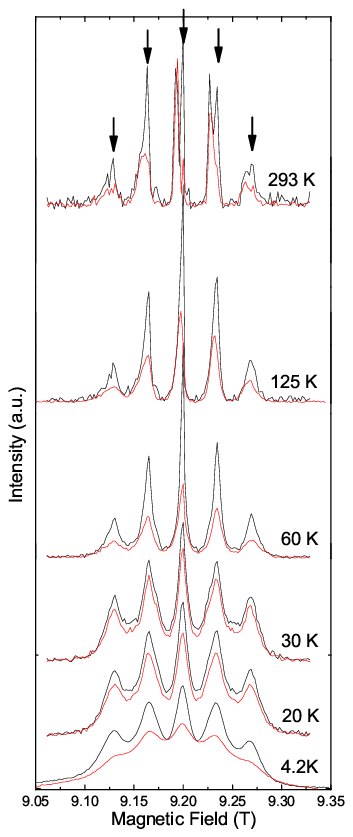}
 \caption{(Color online) Field sweeped \oxy\ NMR spectra of \lbco\ with the external magnetic field parallel to the crystal $c$ axis taken at a fixed frequency of 53.124 MHz. The grey (online red) lines are obtained with a repetition time of 40 ms whereas the black lines are obtained with 500 ms. Therefore, the grey (online red) lines are dominated by the signal of the planar oxygen, whereas the apical signal is suppressed by the fast repetition time. The black lines contain both, apical and planar oxygen. The apical peaks are indicated by arrows. At high temperatures one can clearly distinguish between apical and planar oxygen.}
 \label{Bparac}
\end{center}
\end{figure}

\begin{figure}
\begin{center}
 \includegraphics[width=100mm,clip]{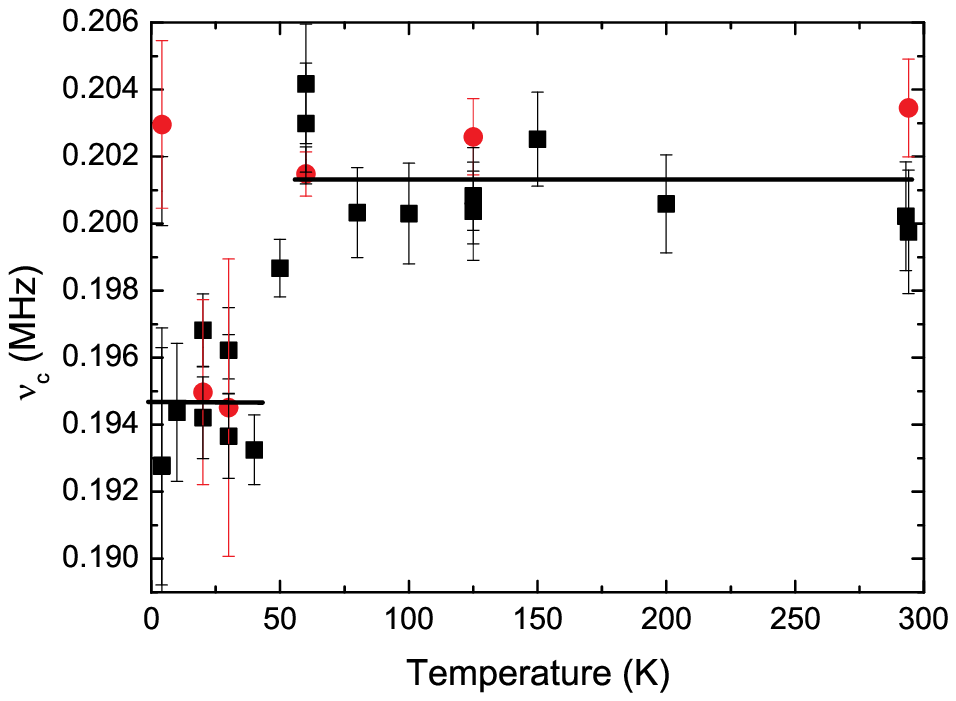}
 \caption{(color online) EFG, $\nu_c$, of the oxygen in \lbco . The black squares are from fits of spectra taken with a fast repetition time, i.e. this is the EFG of the planar oxygen. The grey (color online red) bullets are from fits of the subtracted spectra (see text), i.e. this data shows predominantly the EFG of the apical oxygen. The lines are the average of the black squares for temperatures $\geq$ 60 K and $\leq$ 40 K, respectively.}
 \label{nurnuq}
\end{center}
\end{figure}

There are two oxygen sites in lanthanum cuprates, the planar and the apical sites. The planar site is located within the CuO$_2$ planes directly between two copper atoms. In general, the hyperfine coupling of a nucleus to the electronic spins depends on the wavevector. In the cuprates, the hyperfine field of antiferromagnetic fluctuations cancels at the oxygen, but is maximal for the copper \cite{pines}. The oxygen spin lattice relaxation is therefore damped at the AF wavevector $\mathbf{Q}_{AF}=(\pi/a,\pi/a)$, and measures predominantly the susceptibility at $\mathbf{q}=0$. On the other hand, the apical oxygen is coupled only weakly to the electronic spins of the CuO$_{2}$ planes through the Cu
3d$_{3z^2-r^2}$ orbital, since it is located outside the planes. Therefore, the spin lattice relaxation time of these oxygen sites is much longer than that of the planar ones. This effect has been used to saturate the NMR signal of the apical oxygen by a fast repetition time of the pulse sequence of an NMR experiment \cite{ishida}. Fig. \ref{Bparac} shows temperature dependent field sweeped spectra of the planar and apical oxygen measured with the external magnetic field parallel to the $c$ axis of the single crystal. Two spectra are shown at each temperature, one with a repetition time
of 40 ms, and one with a repetition time of 500 ms. The spectra with the fast repetition rate are dominated by the signal of the planar oxygen, whereas the apical signal is suppressed by the fast repetition. The spectra measured with a long repetition time of 500 ms clearly contain both signals at high temperatures.
At low temperatures, the situation is more complicated, since the spectra broaden. Moreover, the difference in intensity between the spectra at 30 and 20 K measured with long and short repetition times is not as big as at high and very low temperatures (4.2 K), possibly due to a faster relaxation of the apical in this temperature range similar to the $^{139}$La (see below). However, to determine the EFG of the slow and fast relaxing signals, we have subtracted the spectra measured with the fast repetition time from the spectra measured with the long repetition time, and fit the resulting spectra to five Lorentzian lines. This way, we obtain two data sets for the EFG: one from a fit of the spectra with 40 ms repetition time which we attribute to the planar site, and one from a fit of the subtracted spectra, which we attribute to the apical site. The data is shown in Fig. \ref{nurnuq}.

At high temperatures above 60 K, the EFG of the apical oxygen is slightly higher than that of the planar oxygen. Between 60 K and 40 K, the planar EFG drops about 4 $\%$, and stays constant at lower temperatures. The apical EFG also seems to drop below 60 K, though it increases at 4.2 K again. Note that the apical and planar spectra overlap at 30 and 20 K and that it is therefore difficult to clearly distinguish the two. However, the fact that the apical EFG is as high as that of the planar one at high temperatures rules out the scenario in which the measured temperature dependence of the EFG is caused by a complete wipeout of the planar signal. In this case, the apical EFG must be temperature dependent, too, which can be ruled out from the measurements in LESCO (see below).

\subsection{\oxy\ NMR intensity, EFG, and \cu\ NQR intensity}

\begin{figure}
\begin{center}
 \includegraphics[width=90mm,clip]{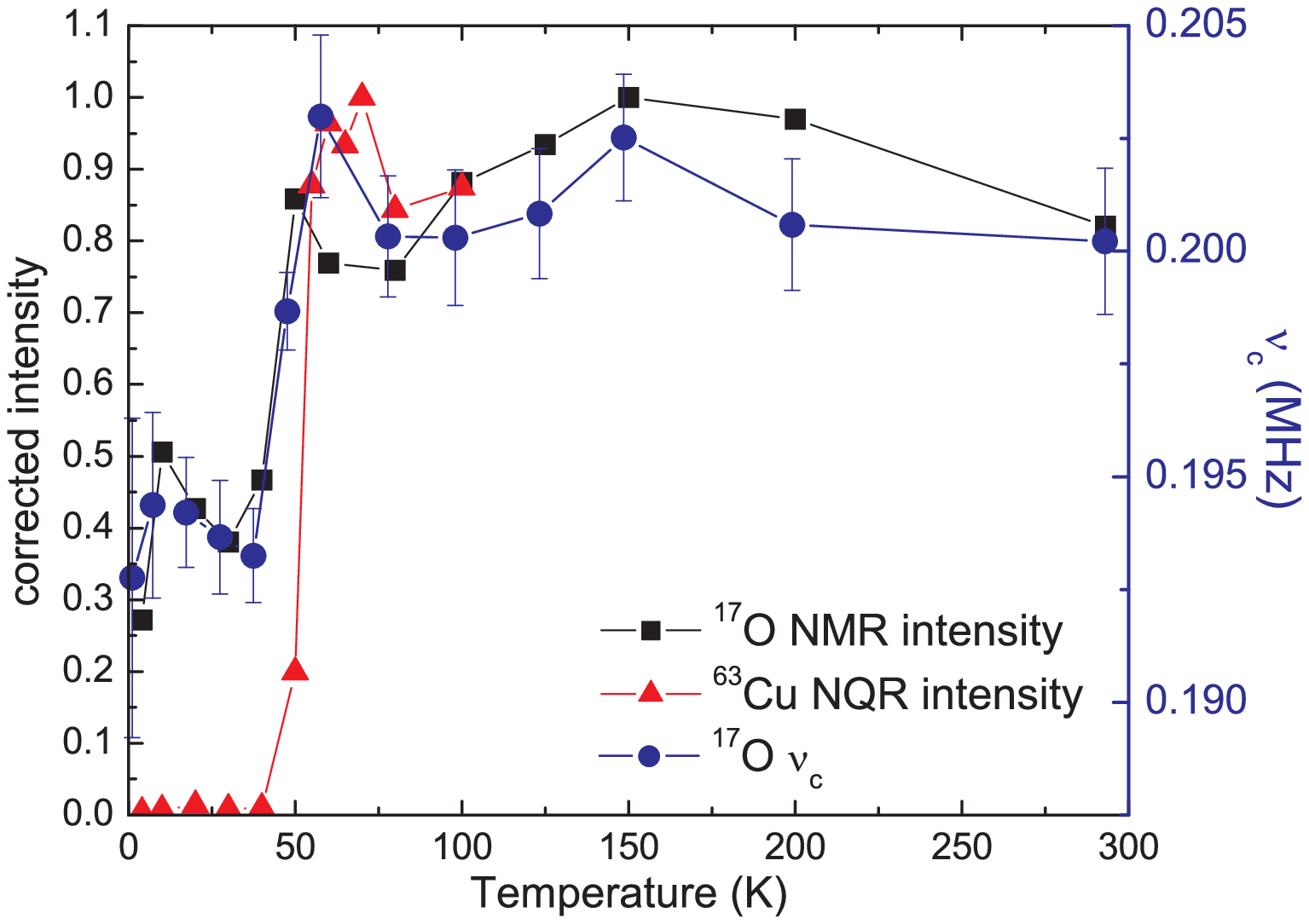}
 \caption{(color online) Integrated and corrected intensity of the oxygen spectra and the EFG of the oxygen obtained in \lbco . The wipeout of the $^{63}$Cu NQR intensity is also shown.}
 \label{nuqLBCO}
\end{center}
\end{figure}

We now turn to a comparison of the oxygen EFG with the integrated intensity of the oxygen and the \cu\ NQR spectral intensity. Figure \ref{nuqLBCO} shows the EFG plotted together with the corrected intensities of the oxygen and Cu NQR. In principle, these quantities should be temperature independent, yet between 60 and 40 K they all change strongly. The Cu NQR intensity completely vanishes, and the \oxy\ NMR intensity drops only about
50 $\%$. The EFG drops at the same temperature by approximately 4$\%$. The wipeout of the Cu intensity is due to the slowing down of the Cu electronic spins in the stripe order. When the Cu electronic spins reach the Larmor frequency of the Cu nuclear spins, they induce transitions in the nuclear spin system so that the Cu nuclei relax before they can be measured. This effect is well known and has been widely discussed in the literature \cite{nick,julien,teitelbaum,imai}. The situation at the planar oxygen site is different. Since the oxygen is located in between two Cu atoms, the hyperfine coupling depends on the wavevector, as discussed above. Here, antiferromagnetic fluctuations of the Cu spins cancel at the oxygen. In a simple stripe picture, the oxygens in the antiferromagnetic ordered regions are not affected by the Cu electronic spin fluctuations, and are consequently not wiped out. In the hole-rich stripes, however, the hyperfine field at the oxygen does not cancel, if there is a hole on one of the copper atoms neighboring the oxygen, and a spin on the other one. In this case, the oxygen indeed experiences a slowly fluctuating hyperfine field from the Cu spins, and its signal is wiped out.  Since these missing oxygen sites are exactly those that couple to a larger hole concentration, $n_h$, associated with the charge stripes, the measured EFG of the remaining oxygens will \textit{decrease}. In other words, the only oxygen sites that remain visible are those distant from the charge stripes, in the antiferromagnetic regions. 

Let us finally assume a scenario, where the planar oxygen is subjected to a complete wipeout, and the remaining signal comes only from the apical oxygen. Then the relaxation of the apical oxygen must increase dramatically below 60 K, and the intensity of the apical oxygen must drop by $\sim 50 \%$, too.

\subsection{Comparison to LESCO}

\begin{figure}
\begin{center}
 \includegraphics[width=90mm,clip]{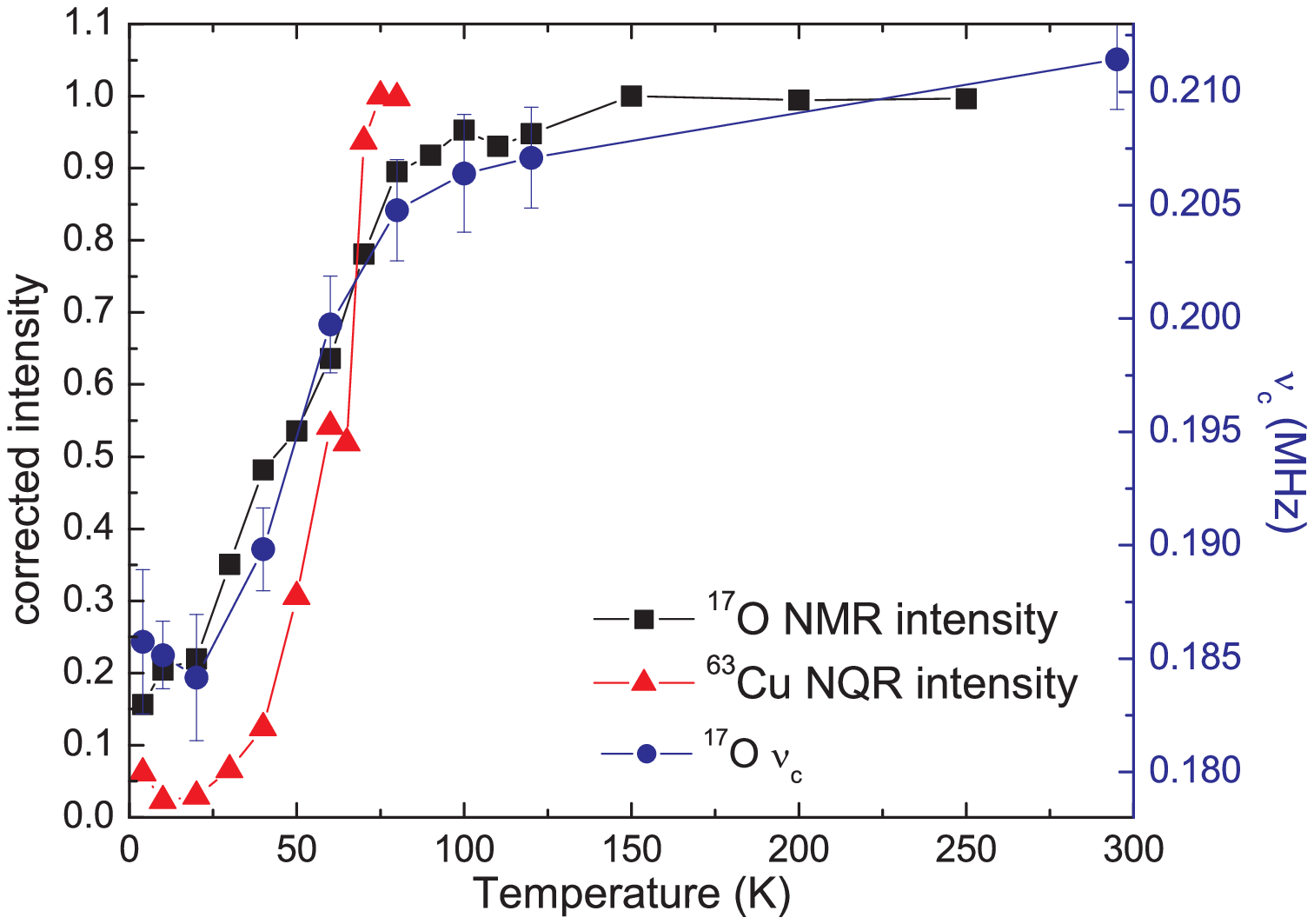}
 \caption{(Color online) Integrated and corrected intensity of the oxygen spectra and the EFG of the oxygen obtained in  ${\rm La}_{1.67} {\rm Eu}_{0.2} {\rm Sr}_{0.13} {\rm Cu O_4}$. The wipeout of the $^{63}$Cu NQR intensity is also shown.}
 \label{nuqLESCO}
\end{center}
\end{figure}

To compare the data obtained in LBCO with LESCO, we have measured the LESCO samples with $x$ = 0.08, $x$ = 0.13, and $x$ = 0.2 again, and confirm our previously published data \cite{grafe}. The new data for $x$ = 0.13 are shown in Fig. \ref{nuqLESCO}.
In comparison to LBCO the drop of $\nu_c$ and of the intensities of the oxygen and the copper develops more smoothly. This may be due to the different phase transition temperatures from LTO to LTT that is 135 K in LESCO and 54 K in LBCO. However, $\nu_c$ and the intensities drop between 80 and 20 K in LESCO with $x$ = 0.13. Surprisingly, $\nu_c$ in LESCO is suppressed by $\sim 12$ $\%$, whereas in LBCO $\nu_c$ is suppressed by only $\sim 4$ $\%$. We do not understand why the absolute value of the change in $\nu_c$ should differ between these compounds, but speculate that the origin of the discrepancy lies in the details of how the charge inhomogeneity is pinned in each system.

\begin{figure}
\begin{center}
 \includegraphics[width=70mm,clip]{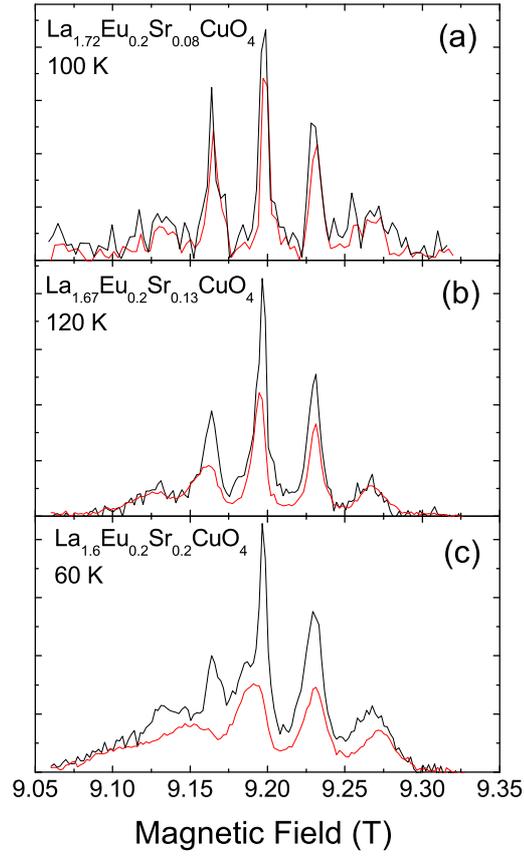}
 \caption{(Color online) (a) to (c): \oxy\ NMR spectra for $x$ = 0.08, $x$ = 0.13, and $x$ = 0.2. The black lines have been measured with a long repetition time of 500 ms, and the grey (online red) lines with a short repetition time of 40 ms.}
 \label{LESCOcomp}
\end{center}
\end{figure}

\begin{figure}
\begin{center}
 \includegraphics[width=80mm,clip]{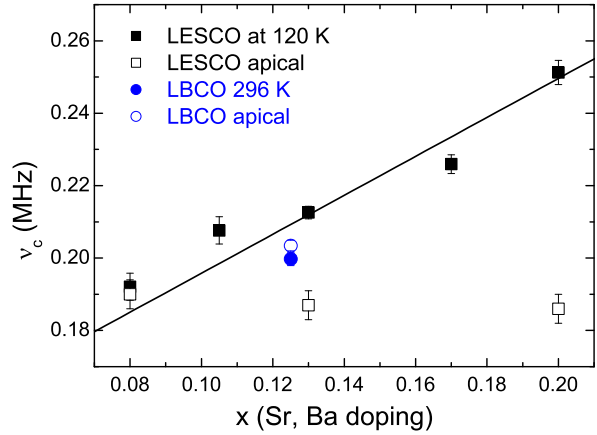}
 \caption{(Color online) Quadrupole frequencies $\nu_c$ for apical and planar oxygen in LESCO and LBCO versus doping $x$ (Sr or Ba). The planar $\nu_c$'s for LESCO are from \cite{grafe}. The solid line is given by $a+bx$, with $a$ = 0.142 MHz, and $b$ = 0.538 MHz.}
 \label{nuqLESCOLBCO}
\end{center}
\end{figure}

Figure \ref{LESCOcomp} (a) to (c) show spectra for $x$ = 0.08, $x$ = 0.13, and $x$ = 0.2 measured with a long and a short repetition time similar to those spectra for LBCO in Fig. \ref{Bparac}. The apical and planar oxygens are clearly distinguishable, especially for $x$=0.20, because for this sample the difference in the quadrupole frequency of the apical and the planar sites is largest. Note, that for the higher dopings there is a substantial asymmetry in the spectra of the planar oxygen that has been observed in LSCO, too \cite{haase2}. This asymmetry comes from a distribution and correlation of the Knight shift and the quadrupole frequency that both depend on the hole doping. In contrast, it seems that the apical oxygen spectra do not exhibit such an asymmetry. Fits of these spectra give the quadrupole frequencies versus Sr and Ba doping that are shown in Fig. \ref{nuqLESCOLBCO}.  
While for LBCO the apical quadrupole frequency is slightly bigger than the planar one, for LESCO the apical quadrupole frequency is always smaller than the planar one. Also, in LESCO the quadrupole frequency of the apical oxygen is not doping dependent in contrast to the planar one \cite{grafe}. Both the planar and apical quadrupole frequencies do not change at the low temperature orthorhombic (LTO) to low temperature tetragonal (LTT) structural phase transition that occurs in LESCO already at $\sim$ 135 K. Therefore, a change of the quadrupole frequency caused by the structural phase transition can also be ruled out. 

\subsection{Spin lattice relaxation of \oxy\ and \la\ in \lbco\ and \lesxco }

\begin{figure}
\begin{center}
 \includegraphics[width=100mm,clip]{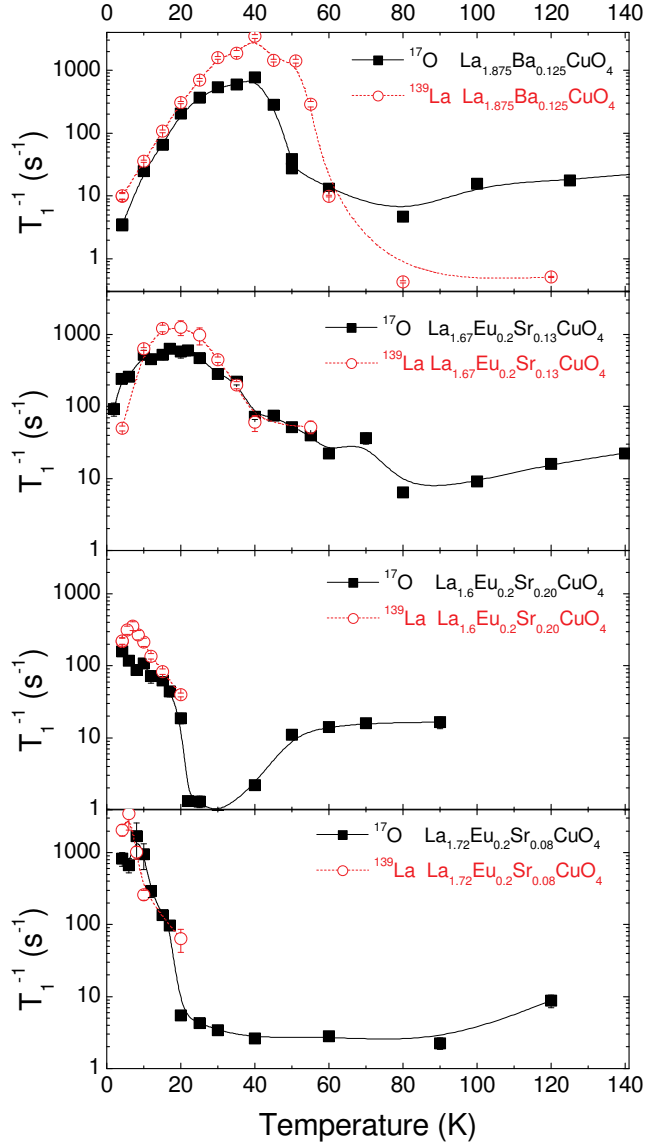}
 \caption{(Color online) Spin lattice relaxation rate $T_1^{-1}$ measured at the planar \oxy\ and the at the \la\ for LBCO and LESCO with
  different Sr contents. The peak occurs in every sample at the same temperature for \oxy\ and \la\ suggesting the same magnetic origin of the peak.}
 \label{T1}
\end{center}
\end{figure}

We have measured the spin lattice relaxation rates for \oxy\ and \la\ in \lbco\ and \lesxco\ for $x$ = 0.08, 0.13, and 0.20 ($x$ = 0.105 and $x$ = 0.17 are not shown). The data are shown in Fig. \ref{T1}. To ensure that we have measured the relaxation of the planar without contributions of the apical, we have measured $T_1$ at the peak of the planar, and integrated only the high frequency part of the Fourier transformed signal. At low temperatures below 50 K we can of course not exclude small contributions of the apical. At high temperatures, the relaxation of the oxygen in  \lesxco\ and \lbco\ is comparable to that in superconducting \lsxco\ for similar doping levels. Most apparent in the data is a broad peak at low temperatures in both the oxygen and the lanthanum relaxation rates. For the lanthanum such a peak has been observed previously \cite{nick,julien,teitelbaum,imai,barbara,borsa,bjsuh,bjsuh2,mitrovic}, and has been related to enhanced magnetic fluctuations of the Cu electronic spins in the stripe ground state, and maybe related to glassy magnetic order. Such a peak occurs, when the correlation time, $\tau$, of the local fluctuating hyperfine field, $h_0$, increases with decreasing temperature, and at some point reaches the Larmor frequency: 1/$\tau = \omega_L$. At this point, the spin lattice relaxation rate given by $1/T_1 = \gamma^2 h_0^2 2\tau/(1+\omega_L^2\tau^2)$ goes through a maximum \cite{purcell}.

The peak in the oxygen relaxation rate occurs at the same temperature as the \la\ $T_1^{-1}$ peak, suggesting that it is induced by the same mechanism. Surprisingly, the magnitude of the peak in the oxygen relaxation rate is lower than that of the lanthanum $T_1^{-1}$. If this peak arises from the apical oxygen, then we would expect a higher relaxation rate than that of the lanthanum since the hyperfine coupling of the apical is greater than that of the lanthanum, and the gyromagnetic ratios of both nuclei are similar ($^{139}\gamma$ = 6.014 MHz/T, and $^{17}\gamma$ = 5.7719 MHz/T). 

\begin{figure}
\begin{center}
 \includegraphics[width=100mm,clip]{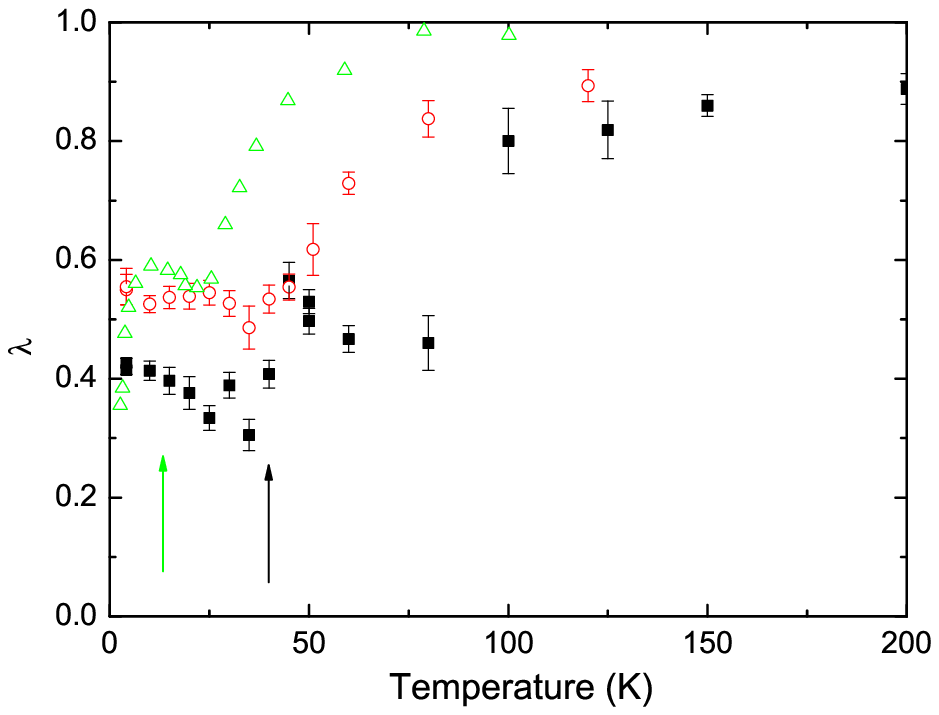}
 \caption{(Color online) Critical exponent, $\lambda$, versus temperature for \lbco\ measured by \oxy\ ($\blacksquare$) and \la\ NMR ($\textcolor{red}{\circ}$). \la\ data for ${\rm La}_{1.88} {\rm Sr}_{0.12} {\rm Cu O_4}$ reproduced from \cite{mitrovic} is indicated by $\textcolor{green}{\triangle}$. The arrows mark the temperature of the maximum of the peak of the spin lattice relaxation rate.}
 \label{lambda}
\end{center}
\end{figure}

However, to compare the different relaxation rates we have to fit the data consistently. For the lanthanum, we have used the standard expression for magnetic relaxation for spin $I$ = 7/2 with a single $T_1$ at high temperatures. In order to be consistent with previous
authors \cite{nick,barbara,bjsuh}, we have fit the data to a stretched exponential form at low temperatures: $M(t)= M_0 \cdot [1-2\exp(t/T_1)^\lambda]$, where $M_0$ is the equilibrium magnetization, and $\lambda$ is the critical exponent \cite{johnston1,johnston2}. Such a stretched exponential form is often used when a distribution of nuclear spin-lattice relaxation rates appears. The critical exponent, $\lambda$ is shown in Fig. \ref{lambda}. The transition temperature from high to low temperatures, i.e. from multi-exponential to stretched exponential, depends in our case on the individual sample and on the doping level: in \lesxco\ for example the increase of \latone\ starts below $T\ \sim $  60 K for
$x$ = 0.13, but only below $T\ \sim $ 25 K for $x$ = 0.08 and $x$ = 0.20. By fitting the data to the stretched exponential form we are able to compare the \la\ \slrr\ to other published data \cite{nick,barbara,bjsuh}.

\begin{figure}
\begin{center}
 \includegraphics[width=90mm,clip]{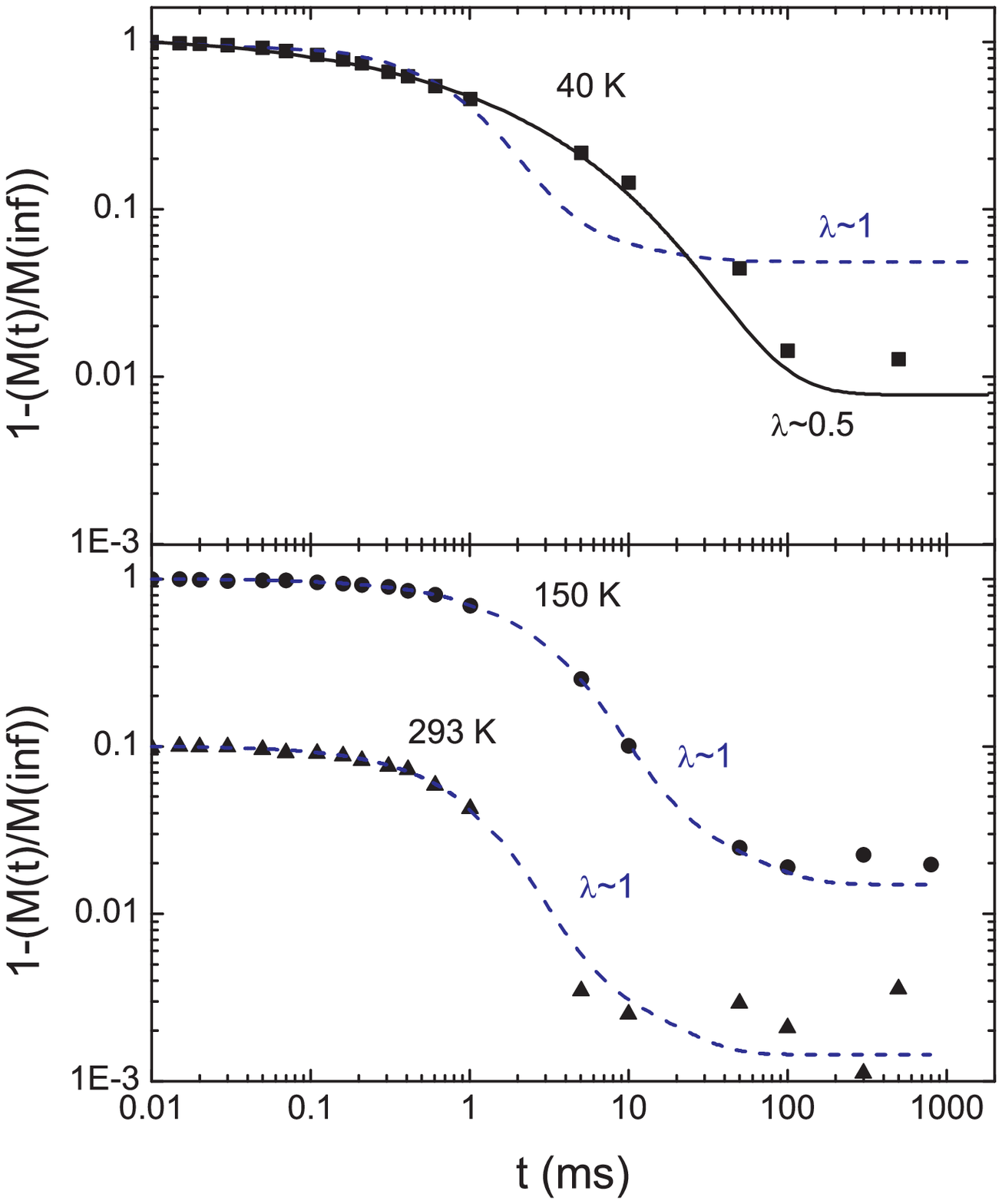}
 \caption{(Color online) $M(t)$ versus $t$ for the oxygen in \lbco\ measured at 293 K, 150 K, and 40 K. The black line is a fit to the stretched exponential, whereas the dashed line is a fit to the standard expression for magnetic relaxation. Clearly, the stretched exponential fits better at 40 K, whereas the standard relaxation function for spin $I$ = 5/2 fits the data at higher temperatures. The data at 293 K is offset by one order of magnitude for clarity.}
 \label{relaxfits}
\end{center}
\end{figure}

\begin{figure}
\begin{center}
 \includegraphics[width=90mm,clip]{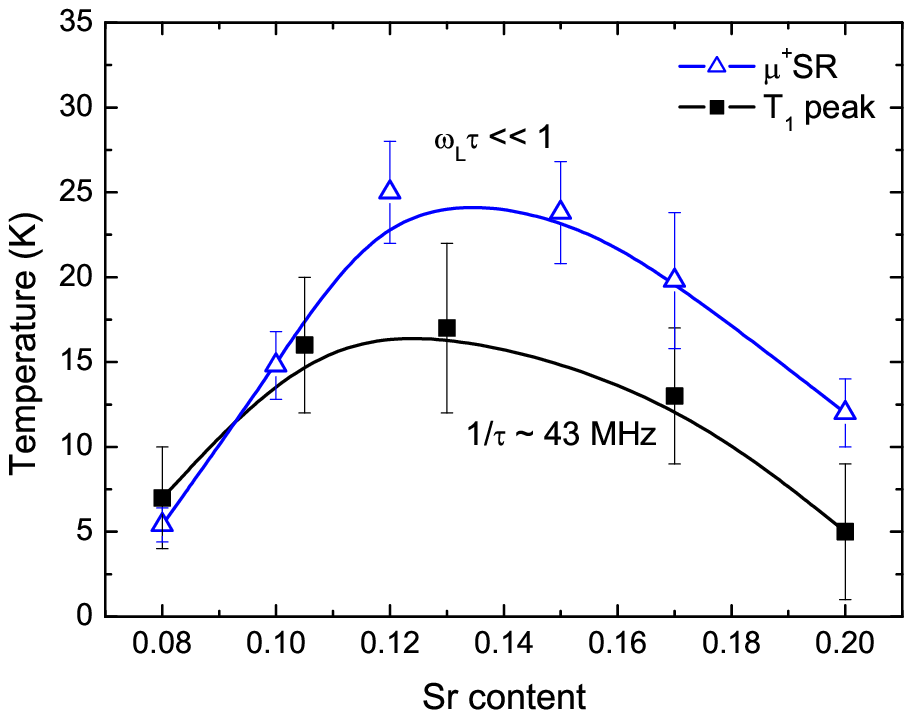}
 \caption{(Color online) Slowing of magnetic fluctuations in \lesxco\ as observed from $\mu$SR \cite{hanshenning} and
 NMR \otone .}
 \label{phasedia}
\end{center}
\end{figure}

For the oxygen we have fitted the high temperature data to the standard expression for magnetic relaxation for spin $I$ = 5/2 with a single $T_1$. This way we can compare our data with superconducting \lsxco\ at these temperatures. At low temperatures we have fitted the data to the same stretched exponential that we used for the \la\ to compare the absolute values of the \la\ and the \oxy . The fits of the magnetization recovery of the oxygen are shown in Fig. \ref{relaxfits} for 293 K, 150 K, and 40 K. At high temperatures the data is well fitted by the standard expression with a single $T_1$. At 40 K only the stretched exponential fit gives a reasonable result. 
The \la\ \slrr\ at the peak at $\sim 40$ K is roughly twice that of the oxygen (Fig. \ref{T1}). We emphasize that we have fitted the low temperature data also to other expressions, but always get higher values for \slrr\ for the lanthanum than for the oxygen. For example we took the multi-exponential magnetic relaxation function, $\phi(t/T_1)$, for the particular nuclear transition for a single $T_1$ component, and made it a stretched function: $\phi((t/T_1)^{\lambda})$. The result is basically the same, i.e. the \latone\ peak is higher than the \otone\ peak, but
the absolute values were lower so that we could not compare our \latone\ data to other published data. The uncertainties of the fits and of the indistinguishable planar and apical oxygen prevent us from a quantitative analysis of the spin lattice relaxation rate. In fact, the peak in the oxygen spin lattice relaxation rate can be assigned tentatively to contributions from the apical oxygen. As shown in Fig. \ref{Bparac} for \lbco\ at around 30 K there is almost no slowly relaxing signal, i.e. the relaxation rate of the apical oxygen is faster at this temperature. This would rather explain the appearance of a peak, since the signal of the planar oxygen is partly wiped out, and for the rest of the planar oxygen the hyperfine field should cancel because of the antiferromagnetic order. On the other hand, as discussed above, the peak in the relaxation rate is smaller for the oxygen than for the lanthanum, though the coupling of the lanthanum should be weaker than the coupling of the planar or apical oxygen. However, at higher doping levels ($x$ = 0.17 and $x$ = 0.20) the peak in \slrr\ is broad and of moderate height and occurs at low temperatures. For $x$ close to optimally doping ($x$ = 0.13 and \lbco ) the peak is still broad, but occurs at higher temperatures and is higher than the peak of the overdoped samples. For the underdoped sample with $x$ =0.105 the peak becomes sharper, occurs at somewhat lower temperatures, and is not as high. And finally for $x$ = 0.08 the peak occurs at even lower temperature and is very sharp, but has gained height and is slightly higher than the peak at
optimally doping. Our results are comparable to other published \la\ relaxation data in \lesxco\ \cite{imai}. From these observations we conclude that (1) the fluctuating hyperfine field, $h_0$, that induces the peak is maximal in the underdoped sample with $x$ = 0.08, and decreases at $x$ =0.105 doping. At $x\approx$ 1/8 doping, $h_0$ increases again before it finally decreases in the overdoped regime. (2) The width of the peak and therefore the distribution of fluctuating hyperfine fields is maximal at around 1/8th doping, suggesting that the inhomogeneous distribution of spins and holes in the copper oxide planes is maximal in this doping regime. At higher doping levels the peak is still quite broad, whereas the width
decreases for lower dopings. (3) As can be seen in Fig. \ref{phasedia} the temperature dependence of the maximum of the peak follows the well known trend that has been obtained by $\mu$SR \cite{hanshenning}.
Note that the position of the peak is frequency dependent \cite{bjsuh,bjsuh2}. We have measured the \la\ relaxation in a field of 7.0494 T, and the oxygen relaxation in a field of 7.444 T for the LESCO samples. This gives a measurement frequency of 42.294 MHz for lanthanum and 43.00 MHz for oxygen. For \lbco\ we have used a magnetic field of 9.2 T for both, lanthanum and oxygen. This gives measurement frequencies of 55.097 MHz and 53.11 MHz, respectively.
The difference in the frequencies of LBCO and LESCO may contribute to the higher temperature of the peak for LBCO. The temperature dependence of the peak that is shown in the phase diagram is also consistent with the pressure dependence of \la\ \slrr\ in LESCO $x$ = 0.15 measured recently \cite{barbara2}. It has been found that the peak in the La relaxation rate decreases in temperature with increasing pressure. The pressure releases the pinning of the stripes and thereby the slowing down of Cu spin fluctuations. Consequently, the peak appears at lower temperatures. In our case,
the development of the stripe order is optimal at $x \approx$ 1/8. At higher doping levels the pinning of the stripes may be released by additional charges and thereby the peak moves to lower temperatures. At lower doping levels the slowing down of spin fluctuations also occurs only at lower temperatures, in agreement with e.g. $\mu$SR \cite{hanshenning}. The phase diagram is also consistent with other published phase diagrams measured by \la\ NQR in \lesxco\ \cite{imai} and neutron scattering \cite{ichikawa}. The phase diagram deduced from the onset of the Cu wipeout  \cite{imai,singer2} deviates from our phase diagram at lower dopings (x $<$ 0.12), where the wipeout of the Cu signal sets in already at higher temperatures. This is maybe due to the different hyperfine couplings of the Cu and the O, and due to the low hole concentration. We do not observe a temperature dependence of the quadrupole frequency for x=0.08 in \lesxco\ nor a wipeout of the oxygen intensity \cite{grafe}.

\section{Conclusion}

The results presented here fit well into the proceeding evidence of inhomogeneities in the CuO$_2$ planes of high temperature superconducting  cuprates. The drop of the EFG in LBCO concomitant with the wipeout of the intensity is compatible with a separation of holes and spins in LBCO similar to LESCO. We could show that the signal of the apical oxygen could be separated from the planar oxygen. The spin lattice relaxation rate of the oxygen and the lanthanum in LBCO and LESCO indicates slow fluctuations of the Cu electronic spins that occur at the highest temperatures close to optimal doping.

We thank J. Haase for helpful discussions. This work has been supported by the DFG through FOR 538 (Grant No. BU887/4). The work at Brookhaven was supported by the Office of Science, U.S. Department of Energy under Contract No. DE-AC02-98CH10886. E.V. acknowledges support from the RFBR grants  08-02-91952-NNIO$\_$a and 07-02-01184-a. The German-Russian cooperation project of the DFG (grant No. 436 RUS 113/936/0-1) is acknowledged. B. Y. acknowledges support from I2CAM and NSC 96-2112-M-009-018-MY2.


\begin{thebibliography}{99}

\bibitem{tranquada1} J.M. Tranquada, B. J. Sternlieb, J. D. Axe, Y. Nakamura, and S. Uchida, Nature \textbf{375}, (1995) 561.

\bibitem{tranquada2} J.M. Tranquada, H. Woo, T. G. Perring, H. Goka, G. D. Gu, G. Xu, M. Fujita, and K. Yamada, Nature \textbf{429}, (2004) 534.

\bibitem{fujita} M. Fujita, H. Goka, K. Yamada, J. M. Tranquada, and L. P. Regnault, Phys. Rev. B \textbf{70}, (2004) 104517.

\bibitem{abbamonte} P. Abbamonte, A. Rusydi, S. Smadici, G. D. Gu, G. A. Sawatzky, and D. L. Feng, Nature Physics \textbf{1}, (2005) 155.

\bibitem{fink} J. Fink, E. Schierle, E. Weschke, J. Geck, D. Hawthorn, H. Wadati, H.-H. Hu, H. A. Durr, N. Wizent, B. B\"uchner, G. A. Sawatzky, Phys. Rev. B \textbf{79}, (2009) 100502.

\bibitem{kim} J. Kim, A. Kagedan, G. D. Gu, C. S. Nelson, and Y.-J. Kim, Phys. Rev. B \textbf{77}, (2008) 180513(R).

\bibitem{kim2} Y.-J. Kim, G. D. Gu, T. Gog, and D. Casa, Phys. Rev. B \textbf{77}, (2008)  064520.

\bibitem{zimmermann} M. von Zimmermann, A. Vigliante, T. Niem\"oller, N. Ichikawa, T. Frello, J. Madsen, P. Wochner, S. Uchida, N. H. Andersen, J. M. Tranquada, D. Gibbs, and J. R. Schneider, Euro. Phys. Lett. \textbf{41}, (1998) 629.

\bibitem{hucker} M. H\"ucker, M. von Zimmermann, M. Debessai, J. S. Schilling, J. M. Tranquada, and G. D. Gu, Phys. Rev. Lett. \textbf{104}, (2010) 057004.

\bibitem{valla} T. Valla, A. V. Fedorov, Jinho Lee, J. C. Davis, and G. D. Gu, Science \textbf{314}, (2006) 1914.

\bibitem{homes} C. C. Homes, S. V. Dordevic, G. D. Gu, Q. Li, T. Valla, and J. M. Tranquada, Phys. Rev. Lett. \textbf{96}, (2006) 257002.

\bibitem{li} Q. Li, M. H\"ucker, G. D. Gu, A. M. Tsvelik, and J. M. Tranquada, Phys. Rev. Lett. \textbf{99}, (2007) 067001.

\bibitem{tranquada3} J. M. Tranquada, G. D. Gu, M. H\"ucker, Q. Jie, H.-J. Kang, R. Klingeler, Q. Li, N. Tristan, J. S. Wen, G. Y. Xu, Z. J. Xu, J. Zhou, and M. v. Zimmermann Phys. Rev. B {\bf 78}, (2008) 174529.

\bibitem{hucker2} M. H\"ucker, G. D. Gu, and J. M. Tranquada, Phys. Rev. B \textbf{78}, (2008)  214507.

\bibitem{singer} P. M. Singer, A. W. Hunt, and T. Imai, Phys. Rev. Lett. \textbf{88}, (2002) 047602.

\bibitem{grafe} H.-J. Grafe, N. J. Curro, M. H\"ucker, and B. B\"uchner, Phys. Rev. Lett. \textbf{96}, (2006) 017002.

\bibitem{haase2} J. Haase, C. P. Slichter, R. Stern, C. T. Milling, and D. G. Hinks, J. Supercond. \textbf{13}, (2000) 723.

\bibitem{singer2} P. M. Singer, T. Imai, F. C. Chou, K. Hirota, M. Takaba, T. Kakeshita, H. Eisaki, and S. Uchida, Phys. Rev. B \textbf{72}, (2005) 014537.

\bibitem{hunt} A. W. Hunt, P. M. Singer, K. R. Thurber, and T. Imai, Phys. Rev. Lett. \textbf{82}, (1999) 4300.

\bibitem{nick} N. J. Curro, P. C. Hammel, B. J. Suh, M. H\"ucker, B. B\"uchner, U. Ammerahl, and A. Revcolevschi, Phys. Rev. Lett. \textbf{85}, (2000) 642.

\bibitem{julien} M. H. Julien, A. Campana, A. Rigamonti, P. Carretta, F. Borsa, P. Kuhns, A. P. Reyes, W. G. Moulton, M. Horvatic, C. Berthier, A. Vietkin and A. Revcolevschi, Phys. Rev. B \textbf{63}, (2001) 144508.

\bibitem{teitelbaum} G. B. Teitel'baum, I. M. Abu-Shiekah, O. Bakharev, H. B. Brom, and J. Zaanen, Phys. Rev. B {\bf 63}, (2000) 020507(R).

\bibitem{imai} A. W. Hunt, P. M. Singer, A. F. Cederstr\"om, and T. Imai, Phys. Rev. B {\bf64}, (2001) 134525.

\bibitem{barbara2} B. Simovi\v{c}, M. Nicklas, P. C. Hammel, M. H\"ucker, B. B\"uchner, and J. D. Thompson, Europhys. Lett. \textbf{66}, (2004) 722.

\bibitem{imai2} P. M. Singer, A. W. Hunt, and T. Imai, arXiv:cond-mat/0302078v1.

\bibitem{gvmw} G.V. M. Williams, J. Haase, M.-S. Park, K. H. Kim, and S.-I. Lee, Phys. Rev. B \textbf{72}, (2005) 212511.

\bibitem{takigawa} M. Takigawa, P. C. Hammel, R. H. Heffner, Z. Fisk, K. C. Ott, and J. D. Thompson,  Phys. Rev. Lett. \textbf{63}, 1865 (1989)

\bibitem{haase} J. Haase, O. P. Sushkov, P. Horsch, and G. V. M. Williams, Phys. Rev. B \textbf{69}, (2004) 094504.

\bibitem{pines} A. J. Millis, H. Monien, D. Pines, Phys. Rev. B \textbf{42}, (1990) 167.

\bibitem{ishida}  K. Ishida, Y. Kitaoka, G.-q. Zheng, and K. Asayama, J. Phys. Soc. Jpn. \textbf{60}, (1991) 3516.

\bibitem{barbara} B. Simovi\v c, P. C. Hammel, M. H\"ucker, B. B\"uchner, and A. Revcolevschi, Phys. Rev. B \textbf{68}, (2003) 012415.

\bibitem{borsa} F. C. Chou, F. Borsa, J. H. Cho, D. C. Johnston, A. Lascialfari, D. R. Torgeson, and J. Ziolo, Phys. Rev. Lett. \textbf{71}, (1993) 2323.

\bibitem{bjsuh} B. J. Suh, P. C. Hammel, Y. Yoshinari, J. D. Thompson, J. L. Sarrao, and Z. Fisk, Phys. Rev. Lett. \textbf{81}, (1998) 2791.

\bibitem{bjsuh2} B. J. Suh, P. C. Hammel, M. H\"ucker, B. B\"uchner, U. Ammerahl, and A. Revcolevschi, Phys. Rev. B \textbf{61}, (2000) R9265.

\bibitem{mitrovic} V. F. Mitrovi\'{c}, M.-H. Julien, C. de Vaulx, M. Horvati\'{c}, C. Berthier, T. Suzuki, and K. Yamada, Phys. Rev. B \textbf{78}, (2008) 014504.

\bibitem{purcell} N. Bloembergen, E. M. Purcell, and R.V. Pound, Phys. Rev. \textbf{73}, 679 (1948).

\bibitem{johnston1} D. C. Johnston, S.-H. Baek, X. Zong, F. Borsa, J. Schmalian, and S. Kondo, Phys. Rev. Lett. \textbf{95}, (2005) 176408.

\bibitem{johnston2} D. C. Johnston, Phys. Rev. B {\bf 74}, (2006) 184430.

\bibitem{hanshenning}  H.-H. Klauss, W. Wagener, M. Hillberg, W. Kopmann, H. Walf, F. J. Litterst, M. H\"ucker, and B. B\"uchner, Phys. Rev. Lett. \textbf{85}, (2000)  4590.

\bibitem{ichikawa} N. Ichikawa, S. Uchida, J. M. Tranquada, T. Niemoeller, P. M. Gehring, S.-H. Lee, and J. R. Schneider, Phys. Rev. Lett.  \textbf{85}, (2000) 1738.

\end{thebibliography}
\end{document}